\documentstyle[aps]{revtex}
\begin{document}
\title{
Connection between low energy effective Hamiltonians
and energy level statistics }
\author{
M.Di Stasio$^{(1)}$ and X. Zotos$^{(1,2)}$
}
\address{
 (1) Institut de Physique Th\'eorique, Universit\'e de Fribourg
 CH 1700 P\'erolles, Fribourg, Switzerland \\
 (2) Institut Romand de Recherche Num\'erique en Physique des
Mat\'eriaux (IRRMA), \\
PHB-Ecublens, CH-1015 Lausanne, Switzerland}
\date{Received\ \ \ \ \ \ \ \ \ \ \ }
\maketitle

\begin{abstract}
We study the level statistics of a non-integrable one dimensional
interacting fermionic system characterized by the GOE distribution.
We calculate numerically on a finite size system the level spacing
distribution $P(s)$ and the Dyson-Mehta $\Delta_3$ correlation.
We observe that its low energy spectrum follows rather the Poisson
distribution, characteristic of an integrable system, consistent with
the fact that the low energy excitations of this system are described by
the Luttinger model.
We propose this  Random Matrix Theory analysis as a probe for the existence
and integrability of low energy
effective Hamiltonians for strongly correlated systems.

\end{abstract}

The idea of studying complex systems by analysis of
the statistical properties of their energy levels goes back to the early
Sixties when Wigner, Dyson, Mehta and others \cite{Wigner,Dyson,Mehta}
proposed a new kind of statistical
mechanics for the spectra of quantum Hamiltonians.

Here one renounces exact knowledge on the nature of the
system but proposes that the coarse statistical properties of the
spectra depend only on the symmetries of the Hamiltonian and not on the
detailed form of the interaction it describes.
This statistical hypothesis then states that the distribution
of $n$ consecutive energy levels of a given system is statistically
equivalent
to the behavior of $n$ consecutive eigenvalues chosen from an ensemble of
random matrices with corresponding symmetries.
The statistical theory of energy levels is the precise mathematical
definition of these ensembles.

Using the language of random matrix theory (RMT) possible ensembles
describing the fluctuations of the eigenvalues are defined \cite{Bohigas,bg}:

when the Hamiltonian is invariant under rotation and time reversal, the
corresponding ensemble is the
GOE (Gaussian orthogonal ensemble, invariant under the orthogonal group);
when time reversal invariance is broken, the GUE
(invariant under the unitary group).
Finally the Poisson distribution corresponds to uncorrelated energy levels.

This theory was first applied in Nuclear Physics and recently received
great interest in studies of quantum billiards connected
to the notion of quantum chaos\cite{bgs}. It is observed that
quantum systems whose classical analogue is fully chaotic give
energy spectra with fluctuations described by a RMT ensemble, while
classically integrable ones exhibit Poisson correlations.

Recently RMT has also been applied in the study of quantum Hamiltonians
describing strongly correlated systems
in the context of Condensed Matter Physics
\cite{Mont}. It was also observed that the level distribution is Poisson
for integrable systems (e.g. by Bethe Ansatz),
while typically changes to GOE for generic many-body
systems\cite{JC,Poil}.
This result emerges from the statistical analysis of
energy levels obtained by exact numerical diagonalization of the
Hamiltonian matrix for a small cluster; it can be used as a numerical test
of integrability.

In this work we propose that an analysis of the distribution of
{\it low lying energy levels} can provide information about the existence of
an integrable effective Hamiltonian describing the low energy physics of the
system. By integrable we mean that there exist an infinite
number of conservation laws (hidden) as in Bethe ansatz systems or
(obvious ones) as in a one-particle type Hamiltonian
(e.g. Fermi or Luttinger liquid\cite{Benoit}).

It is a new tool to extract more information from exact diagonalization of
small systems.

In order to test this idea we have studied a well known example of a
quantum many-body system: spinless fermions in one dimension with nearest
($V_1$) and next nearest neighbor interaction ($V_2$) described by the
Hamiltonian:

\begin{equation}
H = -t \sum_i (c^{\dagger}_i c_{i+1} + h.c.) + V_1 \sum_i n_i n_{i+1}
+ V_2 \sum_i n_i n_{i+2}
\nonumber
\end{equation}
where $c^{\dagger}_i (c_i)$ creates (annihilates) a spinless fermion
at site $i$ (running over an N site lattice with periodic
boundary conditions) and $t$ is the hopping matrix element.

For $V_2$=0 this model (equivalent to the anisotropic Heisenberg model)
is integrable using the Bethe ansatz method and as it was
previously shown \cite{Poil} its level statistics is Poisson like.

Introducing a $V_2$ interaction the model is no more integrable.
The low energy effective Hamiltonian in the weak coupling limit and out of
half-filling is the Luttinger model Hamiltonian as found by
linearizing the spectrum around the two Fermi points \cite{Solyom}.
It is exactly soluble using bosonization, the elementary excitations
being density fluctuations. Actually the Luttinger Hamiltonian
describes the low energy physics of a larger class of one
dimensional interacting systems (in the metallic phase) coined
Luttinger liquids by Haldane\cite{Haldane}.

Therefore we expect that for $V_1, V_2 \neq 0$
the level distribution in the high energy regime will correspond to the
GOE ensemble (the non-integrable case) while in the low energy part of the
spectrum we expect a
deviation from the GOE distribution towards the Poisson one.
We are assuming that the spectrum generated by filling
non-equidistant single particle levels with independent particles is
characterized by Poisson statistics. Although no rigorous analysis
of this assumption exists yet, we expect a behavior analogous to the
case study of the anharmonic oscillator spectrum analyzed by
Berry and Tabor \cite{BerryT}. Actually our results, as we will show
below, lend support for this assumption.

Considering the numerical limitations we have chosen to diagonalize the
Hamiltonian matrix of a system with N=21 sites and M=7 fermions, a
1/3 filling.
In order to apply the RMT the Hamiltonian matrix must be diagonalised in
a subspace of the total Hilbert space in which no symmetries are left
(energy levels in disconnected subspaces are not correlated).
So using the translational symmetry we block diagonalized the
Hamiltonian in $k$-momentum labeled subspaces thus removing all obvious
symmetries. We so obtain 10 independent subspaces corresponding to
momenta $k$ (in units of ${2\pi \over N}$) $k=1,10$
(we omitted the $k=0$ subspace as it possesses
an extra symmetry under reflection).
We diagonalize the matrices (at most of dimension D=5539) using the
Lanczos technique\cite{CW}.

As the number of levels available for analysis in the
low energy part is rather limited and after having
verified that the results obtained are the same for every $k$-subspace,
we averaged the {\it level distributions } obtained for the different
$k$-subspaces. This corresponds to considering
the independent $k$-subspaces as independent realizations of the system.

Having total momentum $k$ different from zero our matrices are
complex as in Hamiltonians with broken time reversal symmetry ($T$).
However, due to the invariance of our system under
reflection symmetry, $I :R \to -R$ the Hamiltonian is still invariant under
$T \times I$ and this invariance leads again to GOE spectral fluctuations
instead of GUE\cite{RB}.

To characterize the fluctuations of $n$ levels with energies
$\{ E_i \}$ ($i=n_0+1,n_0+n$) starting from level $n_0$,
in a given $k$-subspace, we study
the probability density $P(s)$ of spacings between consecutive ordered
levels and the Dyson-Mehta $\Delta_3$ correlation.
As a standard procedure before analysing the fluctuations
we have ``unfolded" the spectrum. This procedure amounts to removing the
smooth irrelevant part of the density function $\langle N(E)_{av}\rangle$.
In practice we consider the new variables:

\begin{equation}
 \delta_i =  N(E_i) - <N_{av} (E_i)>
\nonumber
\end{equation}
where $N(E_i)$ is the number of levels with energy less than $E_i$ and
$<N_{av} (E_i)>$ is constructed by fitting the $n$ level spectrum with a
second order polynomial.
Given this new set of ordered levels we define the spacing
$s_i  =  \delta_{i+1} - \delta_{i}$.

This variable is then rescaled to correspond to a normalized
probability distribution function with average $<s>=1$.
This allows us a direct comparison with the
ideal Poisson $P(s) = \exp \{ -s \}$ or the so called
GOE distribution function:
$P(s) = { \pi \over 2} s \exp \left\{ - {\pi \over 4} s^2 \right\}$

Notice that the GOE distribution with $P(s)\to 0$ for $s\to 0$,
characterizes the level repulsion present in correlated spectra, while
in the Poisson distribution the largest probability is for $s\to 0$
corresponding to level crossings characteristic of uncorrelated spectra.

Now we will describe the results of our study: first to emphasize
the difference
between an integrable and a non-integrable case for our model system
we show in Fig.1 the $P(s)$ for $V_1=2t, V_2=0$ (integrable) and
$V_1=t, V_2=0.5t$ (non-integrable) case. It is clear that indeed the first
follows a Poisson while the second follows a GOE distribution. In the
inset we also show the $N(E_i)$ for $V_1=t, V_2=0.5t$ as a guide for
our further choice of high and low energy windows where we will
perform the partial analysis of the distribution function.

In Fig.2 we present $P(s)$ as estimated from two different energy windows,
the one centered at the low energy part of the spectrum ($n_0=10$),
the other
at the middle part (see inset Fig.1 ): we observe a definite
displacement towards the Poisson distribution for the low energy window,
although the fluctuations in the estimation of $P(s)$
are considerable due the small number of levels used.
The deviation from the
GOE distribution is manifested by an increased weight of
$P(s)$ for $s\to 0$, characteristic of the absence of level repulsion,
but also by an interesting shift at large $s$ towards the
Poisson distribution tail (seen clearly in Fig.3). This deviation
we attribute to the existence of an integrable low energy effective
Hamiltonian as the Luttinger liquid Hamiltonian proposed for this model
\cite{Haldane}. We should stress that with our finite size system and
limited low energy window we do not observe a genuine Poisson distribution
but only a shift from the GOE one. We cannot really say from these data if the
generic low energy level distribution for a macroscopic system is the Poisson
or some other intermediate distribution.

We can significantly improve on the evaluation of $P(s)$,
as is shown in Fig.3, where
results for $P(s)$ for the same windows are presented
but now averaged over all independent $k$-subspaces ($\langle P(s)\rangle_k$).
These results are generic to our system as we also obtained them
for the N=15 (M=5, D=201) and N=18 (M=6, D=1038) systems.
On the other hand we observed no such deviations on a study of a test
random matrix of finite size, so these results are not due to finite size
or density of states effects.

At this point we should mention that we observed similar deviations from
the GOE distribution at the highest energy part of the spectrum which we can
also attribute to a simple level structure,
characteristic of an effective Hamiltonian description obtained by a
unitary transformation ($\tilde c_l= c_l e^{-i\pi l}$) which maps
$H(V_1,V_2) \to -H(-V_1,-V_2)$.

To further study this smooth transition away from the GOE distribution we
studied the $P(s)$ for a group of levels weighted by a Boltzmann factor
which amounts to introducing a soft cutoff procedure in the window of
levels we are studying. Introducing a fictitious temperature $T$, $P(s)$ is
calculated as: $P(s) \simeq \sum_i e^{-\beta E_i} \delta(s-s_i)$
($\beta=1/T$).
For $T\to\infty$ this cutoff procedure is equivalent to a finite
energy window as before;
the results are shown in Fig.4 for different values of $\beta$.
We find the same trend in the results as before,
the results being qualitatively independent of the procedure used
(note that introducing a Boltzmann weight does not affect a spectrum with
pure GOE or Poisson level distribution).
This method can be used for a consistent comparison of level fluctuations
between different size systems.

Finally another probe of level fluctuations introduced by Dyson and Mehta
\cite{Mehta} is the correlation $\Delta_3$ which we calculate as
described by Bohigas and Giannoni\cite{bg2};
\begin{equation}
\Delta_3={1\over 2L} min_{A,B} \int^L_{-L} [N(E)-AE-B]^2 dE
\end{equation}
In Fig. 5 we present the results
again for energy windows at different parts of the spectrum; the Poisson
distribution takes the value $L/15$ while the asymptotic behavior of the
GOE one is $(\ln L)/\pi^2$. The same behavior as for $P(s)$
characterizes $\Delta_3$,
a similarity to the GOE behavior for high energies and a deviation
towards the Poisson one as the energy is lowered. The results shown are
averaged over $k$ subspaces; similar ones where obtained for every
$k$ subspace although with poorer statistics.

In conclusion we studied the level statistics of
a non-integrable system of interacting
spinless fermions in one dimension: we find that although the
overall spectrum is characterized by the GOE distribution, its
{\it low energy part} exhibits a clear tendency towards
the Poisson distribution
characteristic of an integrable system. We attributed this change to
the existence of an {\it integrable} effective Hamiltonian describing the
low lying excitations.
So far we see two classes of Hamiltonians which will give rise to
uncorrelated levels,
characterized by a Poisson distribution: Bethe ansatz systems possessing a
macroscopic number of conservation laws and single particle Hamiltonians
describing practically independent quasiparticles (which notice, might have
nothing to do with the original bare particles as is the case in the
Luttinger model).
{}From these observations
we propose to use this Random Matrix Theory analysis to probe the existence
and integrability of low energy
effective Hamiltonians for strongly correlated systems.
Unfortunately the study of these
simple level correlations seems too crude a
probe to provide information on the nature of the quasiparticle description.

It is even worth considering in future studies the question if {\it all}
Hamiltonians describing physical systems possess a low energy
quasiparticle description.

\acknowledgments

We would like to thank B. Doucot, H. Kunz, D. Baeriswyl and K. Rezakhanlou
for useful discussions. This work was supported by the Swiss
National Fond Grant No. 2000-039528.93 and the University of Fribourg.

\begin{figure}

\caption{$P(s)$ for $V_1 = 2 t$ and $V_2=0$ (circles),
$V_1=t$ and $V_2=0.5t$ (squares), $k=4$ ; continuous lines are
the ideal Poisson and GOE distributions (in the inset $N(E_i)$ for $V_1=t,
V_2=0.5t$). }
\label{one}

\caption{$P(s)$ for $V_1=t$ and $V_2=0.5t$, $k=4$,
from $n=150$ levels; at low energies,
$n_0=10$ (black dots) and at medium energies (squares). }
\label{two}

\caption{$\langle$P(s)$\rangle _k$ (average over $k$-momenta) for the same
parameters as in Figure 2. }
\label{three}

\caption{$\langle$P(s)$\rangle _k$ for $V_1=t$ and $V_2=0.5t$,
($n_0=10, n=2000$) introducing
a Boltzmann weight with $\beta= 0.001,0.005,0.01 $}
\label{four}

\caption{: $\langle \Delta_3(L)\rangle_k$ for $V_1=t, V_2=0.5t$; $n=150$,
energy windows at $n_0=30$ (triangles), $n_0=200$ (black dots),
$n_0=3000$ (squares). Continuous lines: $ (L-2)/15$ (Poisson),
$ (\ln (L-2))/\pi^2$ (GOE) }
\label{five}

\end{figure}

\end{document}